\documentclass[preprint2]{aastex}

\slugcomment{}

\shorttitle{Tori of PWNe}
\shortauthors{Bamba et al.}

\begin{document}

\title{{\it Chandra} View of Pulsar Wind Nebula Tori}

\author{
Aya Bamba\altaffilmark{1,2},
Koji Mori\altaffilmark{3},
Shinpei Shibata\altaffilmark{4}
}

\altaffiltext{1}{
School of Cosmic Physics, Dublin Institute for Advanced Studies
31 Fitzwilliam Place, Dublin 2,
Ireland
abamba@cp.dias.ie
}

\altaffiltext{2}{
ISAS/JAXA Department of High Energy Astrophysics
3-1-1 Yoshinodai, Sagamihara,
Kanagawa 229-8510, JAPAN
}

\altaffiltext{3}{
Department of Applied Physics, Faculty of Engineering
University of Miyazaki,
1-1 Gakuen Kibana-dai Nishi, Miyazaki, 889-2192, Japan
}

\altaffiltext{4}{
Department of Physics, Yamagata University
1-4-12 Kojirakawa,
Yamagata 990-8560, Japan
}

\begin{abstract}
The results from a systematic study of
eleven pulsar wind nebulae
with a torus structure observed with the {\it Chandra} X-ray observatory
are presented.
A significant observational correlation is found
between the radius of the tori, $r$,
and the spin-down luminosity of the pulsars, $\dot{E}$.
A logarithmic linear fit between the two parameters yields
log\,$r$ = (0.57 $\pm$ 0.22) log\,$\dot{E}$ -22.3 $\pm$ 8.0
with a correlation coefficient of 0.82,
where the units of $r$ and $\dot{E}$ are pc and ergs~s$^{-1}$, respectively.
The value obtained for the $\dot{E}$ dependency of $r$
is consistent with a square root law,
which is theoretically expected.
This is the first observational evidence of this dependency,
and provides a useful tool to estimate the spin-down energies of pulsars
without direct detections of pulsation.
Applications of this dependency to some other samples are also shown.
%
\end{abstract}


\section{Introduction}

Active pulsars eject relativistic pulsar winds comprised of
relativistic particles and magnetic field. Such winds are
terminated by a strong shock where pressure balance is attained with
the ambient medium. High energy particles diffusing out in the
downstream of the shock emit radio to very high energy gamma-rays via
synchrotron or inverse Compton processes, which are observed as pulsar
wind nebulae (PWNe).  The distance from the pulsar to the termination
shock, $r_{\rm s}$, is expected to be
\begin{equation}
r_{\rm s} = \left(\frac{\dot{E}}{4\pi c\eta P_{\rm ext}}\right)^{1/2}\ \ ,
\label{eq:r_s_riron}
\end{equation}
where $\dot{E}$, $c$, $\eta$, $P_{\rm ext}$ are the spin-down luminosity
of the pulsar, the light speed, the filling factor, and the external
pressure \citep{rees1974}.  The size of shocks should be roughly
0.01--0.1~pc with typical parameters of PWNe \citep{kennel1984}.  
In the case of a weakly-magnetized pulsar wind, which is believed to
apply to most of these systems, particle energy density dominates that of
magnetic field energy at the shock in which the emissivity of
synchrotron radiation is relatively low. Then, the shock is generally
invisible with the current instruments except for the brightest example,
the Crab Nebula \citep{weisskopf2000}.
The equipartition between
particle and magnetic field energy, at which the Synchrotron emissivity
is the highest, is reached at a distance of a few times of $r_{\rm s}$,
as the post shock flow decelerates
\citep{kennel1984}.
Considering that most of the pulsar wind energy is
blown near the equatorial plane and that Synchrotron cooling becomes
efficient at the outer region, a torus-like structure is expected in this
system whose radius is a few times of $r_{\rm s}$. 


The {\it Chandra} X-ray observatory has actually revealed the torus (and
jet-like) structures from several PWNe thanks to the excellent spatial
resolution of 0.5~arcsec \citep[c.f.,][]{kargaltsev2008}.
\citet{ng2004} developed a sophisticated method
to fit three-dimensional model to the tori,
which they applied to the Chandra data of
several PWNe \citep{ng2008},
providing most reliable values of the torus radius.
Since the termination shock radius is expected to be
proportional to the square root of the spin-down luminosity,
the torus radii may scale in the same way as the shock radii,
if so, one would expect a correlation between the torus radii and the
spin-down luminosities following the square-root law.
In this paper, we confirm this relationship for the first
time, and show its astrophysical use for some PWNe.

\section{Samples}

In this paper, we used 11 samples for our study
with the following criteria.
{\it Chandra} observed several tens of PWNe \citep{kargaltsev2008}.
Thanks to the excellent spatial resolution of the X-ray telescope,
equatorial tori and polar jet structures 
are discovered in more than 10 PWN systems.
\citet{ng2004,ng2008} measured the torus radius of 10 PWNe
with the developed method by \citet{ng2004}.
\citet{romani2005} uses the same method to B1706$-$44,
and we added this PWN to our sample.

Table~\ref{tab:parameters} shows details of our samples.
Our samples are so young that the pulsar wind is still strong and
morphological distortion due to pulsar motion is relatively small yet;
old systems sometimes show cometary structure controlled by this effect.
Actually, they are categorized to ``PWNe with toroidal components''
in \citet{kargaltsev2008}
except for J0537$-$6910,
which has a cometary nebula \citep{chen2006}.
\citet{ng2008} derived the size of its torus
($r_{\rm arcsec}$ in Table~\ref{tab:parameters})
after subtracting the diffuse nebula,
and we use the parameters by \citet{ng2008}.
The latest distance estimates are used
to calculate the physical radius of the tori
($r$ in Table~\ref{tab:parameters})
with references in Table~\ref{tab:parameters}.

\section{Results}

\begin{figure}
\epsscale{0.9}
\plotone{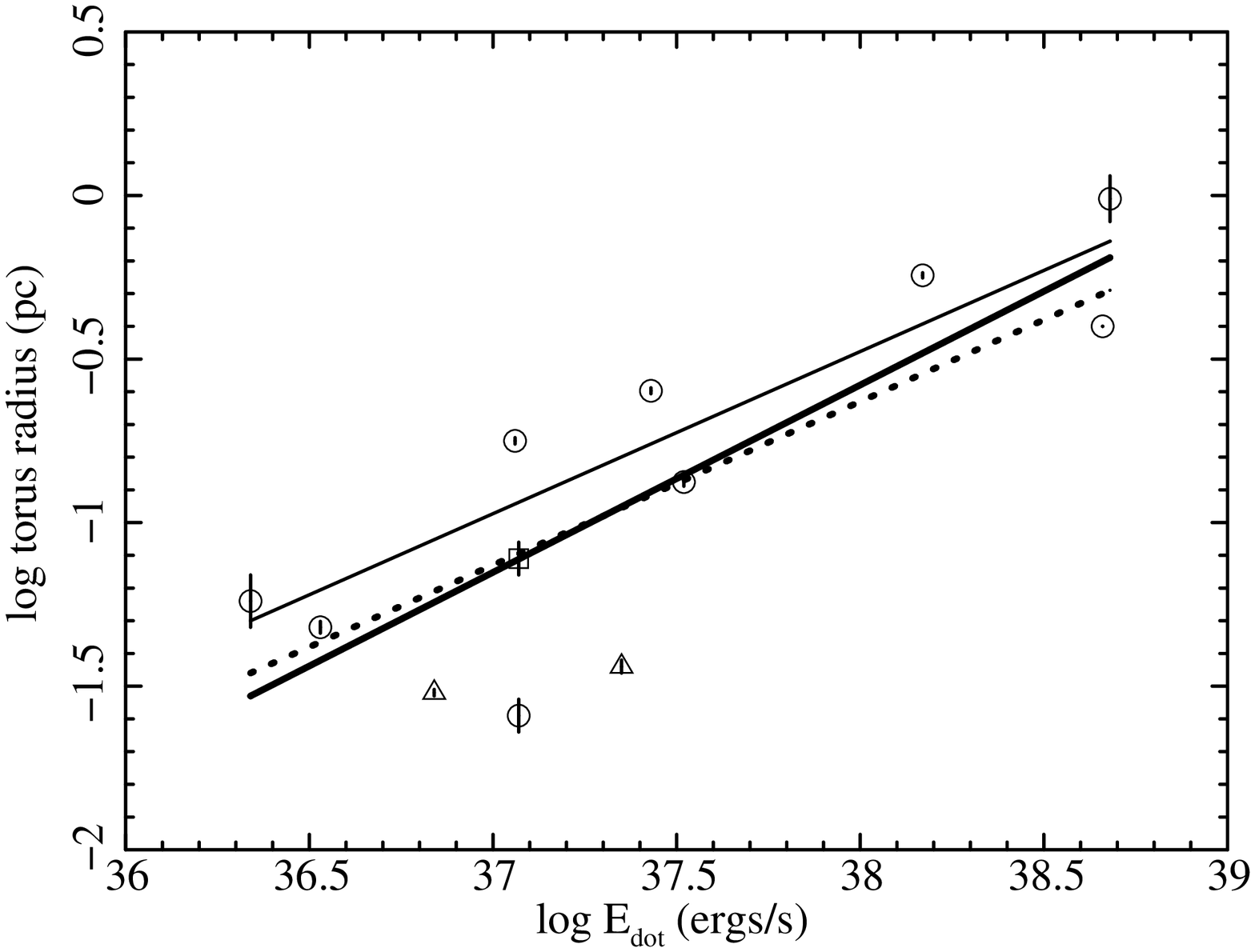}
\caption{Spin-down luminosities vs. radii of tori.  Triangles and
circles are samples with/without interaction with interstellar medium or
ejecta, while the box is a new estimation of the radius for
PSR~J1124$-$5916 (see text). Errors are quoted from
Table~\ref{tab:parameters}.  Thick solid and dashed lines represent
eq.(\ref{eq:result2}) and (\ref{eq:result}), whereas the thin solid line
is for eq.(\ref{eq:9samples}).  }
\label{fig:Edot-r}
\end{figure}

In this section, we search for a correlation 
between the torus radius $r$ and spin-down luminosity $\dot{E}$. 
Figure~\ref{fig:Edot-r} represents the plot of
$r$ as a function of $\dot{E}$.
Uncertainties of $r$ in the figure are obtained simply by multiplication
of the nebula distances and the statistical uncertainties in apparent
torus radii determined by \citet{ng2004,ng2008} and \citet{romani2005}.
It is apparent that there is a strong positive correlation between $\log
\dot{E}$ and $\log r$, yielding a correlation coefficient of 0.82. This
power-law like correlation is as expected as we review in \S~1.
On the
other hand, it is also obvious that there is a non-negligible
fluctuation beyond a simple power-law function owing to the
statistical uncertainties. These facts suggest that $r$ is certainly a
function of $\dot{E}$ but that there are other hidden parameters which
give a larger fluctuation to this relation than the given statistical
uncertainties. We will discuss possible origin of the parameters
later. We thus fit the data weighting them equally with a power-law function,
\begin{equation}
\log r = \alpha \log \dot{E} + \beta\ \ ,
\end{equation}
where $\alpha$ and $\beta$ are constant values. We obtained
\begin{eqnarray}
\alpha &=& 0.57\pm 0.22 \ \ ,
\nonumber \\
\beta &=& -22.3\pm 8.0 \ \ ,
\label{eq:result2}
\end{eqnarray}
respectively.
The best-fit model is shown in Figure~\ref{fig:Edot-r}
with a thick solid line.

\begin{figure}
\epsscale{0.6}
\plotone{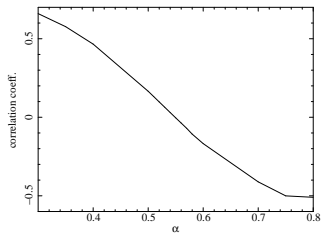}
\caption{Correlation factors with different $\alpha$.}
\label{fig:corr}
\end{figure}

In order to check our result from a different point view,
we calculated the correlation coefficient between $\log \dot{E}$ and
$r/\dot{E}^\alpha$,
which should be 0 with the best-fit $\alpha$.
Figure~\ref{fig:corr} shows
the relation between $\alpha$ and the correlation coefficient.
We can see that
the correlation coefficient becomes close to 0
when $\alpha$ is between 0.5 and 0.6.
This result indicates again that
$\alpha$ is around 0.5--0.6.

The value of $\alpha$ agrees well with 
the expected 0.5 (eq.(\ref{eq:r_s_riron})).
It also implies that the torus radius scales
in the same manner as the radius of the termination shock.
If we fix $\alpha$ to be 0.5, the equation becomes
\begin{eqnarray}
\log r &=& 0.5 \log\dot{E} -19.6\pm0.2\ \ 
\label{eq:result}.
\end{eqnarray}
This result is also shown in Figure~\ref{fig:Edot-r}
with a thick dashed line.

There are three samples which are well below the best-fit lines:
Vela, PSR~J2229+6114, and PSR~J1124$-$5916. The former two are known to
show clear evidences of interaction with ejecta or interstellar medium
\citep{lamassa2008,kothes2001}.%
In such a case, as is indicated by
Eq.(\ref{eq:r_s_riron}), the torus radii could be smaller confined by
higher external pressure compared to those without interaction with
surrounding medium. As for the last one, Park et al. (2004) showed that
the reverse shock has not yet begun to interact with it. However, the
latest 510~ksec Chandra observation revealed the almost 3 times larger
torus compared with that seen in the previous 50~ksec
Chandra observation which Ng \& Romani (2008a) analyzed (Park et al. 2007).

Taking these facts into account, we made another fit where Vela and
PSR~J2229+6114 are excluded and the original value of PSR~J1124$-$5916
is multiplied by a factor of 3. In this fit, we found the better
correlation factor of 0.93 and obtained the best-fit model of
\begin{equation}
\log r = (0.49\pm0.12) \log \dot{E} - 19.2\pm4.7
\label{eq:9samples}
\end{equation}
as shown with a thin solid line in Figure~\ref{fig:Edot-r}.

We also searched for correlation between
the torus radius and other physical parameters,
such as age of PWNe, magnetic field, and so on,
but could not find any significant correlation.

\section{Discussion}

\subsection{Termination shocks and tori}

Eq.(\ref{eq:result2}) shows that
the tori radii ($r$) show a square root dependance on $\dot{E}$
like the termination shock radii ($r_{\rm s}$):
This is the first clue of the termination shocks in PWNe.

Here, we introduce $z \equiv r/r_{\rm s}$,
the ratio between the radii of observed torus and the shock.
The value of $z$ can be common among pulsars in spite of
different pulsar parameters and environment.
Assuming eq.(\ref{eq:result}),
\begin{eqnarray}
z &=& r/r_{\rm s} \nonumber \\
&=& 10^\beta(4\pi c\eta P_{ext})^{1/2} \nonumber \\ 
&=& 1.9_{-0.7}^{+1.1}\left(\frac{\eta}{1}\right)^{1/2}\left(\frac{P_{ext}}{1.6\times 10^{-9} {\rm [g~cm^{-1}s^{-2}}]}\right)^{1/2} \ \ .
\end{eqnarray}
When the temperature, number density, and filling factor of 
external environment of the termination shock is
1~keV, 1~cm$^{-3}$, and 1,
($P_{ext} = 1.6\times 10^{-9}$~g~cm$^{-1}$s$^{-2}$),
the ratio of radii of X-ray torus and termination shock is
almost unity.
The X-ray emission should come from the outside region of the termination shock
according to \citet{kennel1984},
which supports our result.

The best-fit $\alpha$, 0.57, is slightly larger than
the theoretical value in eq.(\ref{eq:r_s_riron}).
The data scatter seems to be larger at lower values for spin-down energy.
It could be due to an age effect.
The fitting without Vela and J2229+6114,
which interact with ejecta or interstellar medium,
shows larger radius and smaller dispersion.
This result supports our scenario.


\subsection{Fluctuation of thermal parameters}


Figure~\ref{fig:Edot-r} shows that the data points have rather
large scatter to the best-fit line.
We here discuss on the possible origin of the fluctuation.
Since $\dot{E}$ is derived solely from timing information
which generally has much higher accuracy than other observable
quantities,
we concentrate on other parameters,
such as thermal parameters like $r$, $\eta$ and $P_{ext}$,
and distance.

We first estimate how large fluctuation of $r$ is required to reproduce
the correlation coefficient of 0.82 by a simple simulation.
A single trial of the simulation generates 11 samples from a parent
population with fluctuated environment, which is
simulated by log-normal variation in the product $\eta P_{ext}$
with a given standard deviation of $\sigma_0$.
We made 1000 trials with a given $\sigma_0$ to calculate
an expected correlation coefficient $<A>$ and the probability
for $A$ to be larger than 0.82.
Table~\ref{tab:fluctuation} lists the expected correlation coefficient and the
probability with different $\sigma_0$'s.
The simulation suggest that the observed scatter is
explained if $\sigma_0 \sim 0.6$, i.e.,
if fluctuation in $\eta P_{ext}$ is a factor of $\sim 4$.

One may think that the distance uncertainty may be a primary source of
this fluctuation.
If it is the case,
we need a factor 2 fluctuation of distance
since the fluctuation of the torus radii is linearly connected to
the distance uncertainty,
which is the square root of $\eta P_{ext}$ fluctuation.
Although distance to astronomical bodies is not
always constrained very well, the factor of 2 appears to be too large to
be attributed to the distance uncertainty alone
especially for such famous and well-studied samples.
One of the most famous
and general measurement method of distance to supernova remnants (SNRs)
is to use the $\Sigma$---$D$ relation
\citep[e.g.,][]{case1998};
the distances to more than 200 SNRs are estimated
using the surface brightness at 1~GHz and diameter
relation,
although there is $\sim$40\% dispersion between distance from their
method and those from other methods.
This is because $\Sigma$---$D$ relations can be used to estimate
properties of ensembles of SNRs,
not for individual one,
as \citet{case1998} mentioned.
Our relation on the PWN tori can be used
in similar way to $\Sigma$---$D$ relation
to estimate their distance.

The fluctuations of $\eta$ and $P_{\rm ext}$ result in that of $r_{\rm s}$, 
and thus that of $r$ as well.
A fluctuation of $r$ by a factor of 2
roughly corresponds to that of $\eta P_{\rm ext}$ by a factor of 4.
This level of fluctuation could easily occur about $P_{\rm ext}$.  The
density of external gas, which is inside of SNR shells, differ by about
three orders of magnitude from SNR to SNR, from $\sim 0.1$
\citep[SN~1006;][]{yamaguchi2008}
up to $\sim 200$ \citep[Cas~A; ][]{lazendic2006}.
Thus, we believe that the variation of $P_{\rm ext}$ is the primary
cause of the scatter.
Actually, the torus of Vela X, which evidently interacts with ejecta,
has smaller radius relative to the best fit function, which may be due
to the larger $P_{\rm ext}$ effect among our samples.
Samples without these PWNe shows clearly tighter correlation.
The fluctuation of $\eta P_{\rm ext}$ is also estimated
with eq.(\ref{eq:9samples})
to be 2.5,
as shown in Table~\ref{tab:fluctuation2},
which is much smaller than those with all samples.

The interesting thing is that all of the samples
with small torii are categorized into combined type SNRs,
which have radio shell and PWNe.
It could be due to that SNRs with bright shells have higher densities
inside the remnants.
The detailed
systematic and observational study of $P_{\rm ext}$ for individual
objects might give a better explanation of this fluctuation although it
is beyond the scope of this paper.


All we mention above are the ``negative'' factors which tend to wash out
the correlation between $\dot{E}$ and $r$. Nonetheless, we find it with
relatively high significance, which in turn suggests the robustness of
this correlation.

\section{Application}

Once the correlation between $\dot{E}$ and $r$ is established, it will
provide a useful tool to estimate the spin-down energy of pulsars without
direct detections of pulsation,
although the large error range prevent us from 
precise parameters determinations.
We show some examples of the application.
In this section, we cite eq.(\ref{eq:result2})
although using eq.(5) does not alter the results here so much.

\subsection{$\dot{E}$ determination for G0.9+0.1}

With eq.(\ref{eq:result2}),
we can estimate the spin-down luminosity of PWNe
from the radius of tori,
without information of pulsation.

Let us consider the example of G0.9+0.1.
The SNR G0.9+0.1 has a X-ray bright PWN in the Galactic center region.
\citet{gaensler2001} resolved the PWN with Chandra.
The size and flux are 5~arcsec $\times$ 8~arcsec
and $6.6\times 10^{-12}$~ergs~cm$^{-2}$s$^{-1}$
in the 2--10~keV band, respectively.
The torus radius should be about half of the longer extension,
4~arcsec,
or 0.16~pc with the distance of 8.5~kpc.
With this value and eq.(\ref{eq:result2},
we can estimate
the $\log \dot{E}$ of G0.9+0.1 to be 37.7.
There is no difference whether we use eq.(\ref{eq:result})
or eq.(\ref{eq:result2}).

\citet{possenti2002} discovered the empirical relation between
$\dot{E}$ and the 2--10~keV luminosity $L_X$ (ergs~s$^{-1}$) of PWNe of
\begin{equation}
\log L_X = 1.34\log \dot{E} - 15.34\ \ ,
\label{eq:possenti}
\end{equation}
although the dispersion is large \citep[c.f.][]{kargaltsev2008}.
We can estimate the spin-down luminosity of G0.9+0.1 independently
from their relation
to be $\log\dot{E}$ = 37.4,
which shows very good agreement with our estimate.

Very recently, \citet{camilo2009} discovered coherent pulsation
from the central pulsar
with the period and the period derivative of 52~ms and
$1.5557\times 10^{-13}$~s~s$^{-1}$, respectively.
The resultant spin-down energy is $\log \dot{E}$ of 37.6,
which also shows good agreement with our result.

\subsection{$\dot{E}$ determination for G328.4+0.2}

\citet{gelfand2007} found an extended structure in the PWN G328.4+0.2
with {\it XMM-Newton} with the size of $\sim 1$~arcsec,
or 0.09~pc at 17~kpc \citep{gelfand2007}.
The coherent pulsation has not been detected yet.
We can estimate the spin-down luminosity in the same way as for G0.9+0.1
to be $\log \dot{E} = 37.3$.
This is totally consistent with previous estimation
using \citet{possenti2002}, $\log \dot{E} = 37.2$ \citep{gelfand2007}.

\subsection{Distance determination of PSR~J1846$-$0258}

PSR~J1846$-$0258 in SNR Kes~75 is one of the mysterious PWNe
with soft-Gamma-ray repeater like flares
\citep{gavriil2008,kumar2008}.
The spin-down luminosity is rather large,
$\log\dot{E}$ of 36.91.
However, the distance of this interesting source is still unclear:
\citet{becker1984} estimated that the distance is 19--21~kpc,
whereas \citet{leahy2008} suggests that this system is much nearer,
5.1--7.5~kpc.

Eq.(\ref{eq:result2})
and the spin-down luminosity indicate
a torus radius of 0.07~pc,
whereas the detected torus by \citet{ng2008b}
has the radius of 10~arcsec.
It suggests that the expected distance from $\dot{E}$---$r$ relation
is about 1.1~kpc.
It is too small compared with other distance estimates,
or in other word, the torus radius is too large.
This would indicate that
either an exceptionally low ambient pressure
or the pulsar provides additional pressure from the inside.
It could be important information to understand the origin of magnetars,
which is still hotly debated
\citep[e.g.,][]{vink2006,ferrario2006,duncan1992,gavriil2008}.
Systematic study of PWN torii of magnetars should be done
although we have few samples until now \citep{rea2009,vink2009}.




\subsection{Torus search of DEM~L241}


The SNR DEM~L241 in the Large Magellanic Cloud (LMC)
has a compact X-ray source in its center 
detected with {\it XMM-Newton} \citep{bamba2006}.
The flux and photon index are $5.0\times 10^{-12}$ in the 2.0--10.0~keV band
and 1.57, respectively.
The spin-down luminosity expected by \citet{possenti2002} is 
$\log E_{dot} = 38.4$,
which is one of the largest values among known PWNe.
We can confirm it when we can detect the torus of the PWNe.
The size estimated to be 0.39~pc, or 1.6~arcsec
using the distance to the LMC of 50~kpc \citep{feast1999},
which can be detectable
with excellent spatial resolution of Chandra,
but not with XMM-Newton \citep{bamba2006}.

\section{Summary}

We have made a systematic study of 
PWN spin down luminosity and tori radii
using {\it Chandra} data of 11 samples.
It is discovered, for the first time, that
$\log r$ and $\log\dot{E}$ have very strong positive correlation.
The tori could be X-ray emission originating
from the outside of the termination shocks
as suggested by \citet{kennel1984}.
The fluctuation in the correlation is mainly producted by
variation of the external pressure and distance uncertainties.
With this correlation, we can estimate
the spin-down luminosity and distance to the PWN
without information of coherent pulsations.
This estimation has been applied to G0.9+0.1, G328.4+0.2,
PSR~1846$-$0258, and DEM~L241.

\acknowledgements

The authors thank the anonymous referee
for his/her suggestions.
The authors also thank T.~Dotani and Y.~Terada
for their comments.
AB is supported by JSPS Research Fellowship for Young Scientists.
The work of KM is partially supported by the Grant-in-Aid for Young
Scientists (B) of the MEXT (No.\ 18740108).


\begin{deluxetable}{p{10pc}ccccccc}
\tabletypesize{\scriptsize}
\tablecaption{Parameters of Tori of PWNe
\label{tab:parameters}}
\tablewidth{0pt}
\tablehead{
\colhead{} & \colhead{radius ($r_{\rm arcsec}$)} & \colhead{distance} & \colhead{radius ($r$)} & \colhead{log$\dot{E}$} & \colhead{log$\tau$} & \colhead{References} \\
 & [arcsec] & [kpc] & [pc] & [ergs~s$^{-1}$] & [years]
}
\startdata
PSR~J0537$-$6910 (N157B)\dotfill & 4.0$\pm$0.58 & 50 & 0.97$\pm$0.14 & 38.68 & 3.70 & (1) (4) \\
Crab\tablenotemark{a}\dotfill & 41.33$\pm$0.20 & 2 & 0.400$\pm$0.002 & 38.66 & 3.09 & (2) (5) \\
PSR~B0540$-$69\dotfill & 2.35$\pm$0.06 & 50 & 0.57$\pm$0.01 & 38.17 & 3.22 & (1) (4) \\
PSR~J1833$-$1034 (G21.5$-$0.9)\dotfill & 5.7$\pm$0.2 & 4.8 & 0.133$\pm$0.004 & 37.52 & 3.69 & (1) (6) \\
PSR~J0205+6449 (3C58) \tablenotemark{a}\dotfill & 16.3$\pm$0.3 & 3.2 & 0.253$\pm$0.005 & 37.43 & 3.73 & (1) (7) \\
PSR~J2229+6114\dotfill & 9.3$\pm$0.4 & 0.8 & 0.036$\pm$0.001 & 37.35 & 4.02 & (2) (8) \\
PSR~J1124$-$5916\dotfill & 0.9$\pm$0.1 & 6 & 0.026$\pm$0.003 & 37.07 & 3.46 & (1) (9) \\
SNR G54.1+0.3\dotfill & 4.6$\pm$0.1 & 8 & 0.178$\pm$0.004 & 37.06 & 3.46 & (2) (10) \\
Vela\dotfill & 21.25$\pm$0.50 & 0.29 & 0.030$\pm$0.001 & 36.86 & 4.05 & (2) (11) \\
PSR~B1706$-$44\dotfill & 3.30$\pm$0.13 & 3 & 0.048$\pm$0.002 & 36.53 & 4.23 & (3) (12) \\
PSR~B1800$-$21\dotfill & 3.1$\pm$0.5 & 3.8 & 0.057$\pm$0.009 & 36.34 & 3.21 & (1) (13) 
\enddata
\tablecomments{(1) \citet{ng2008}; (2) \citet{ng2004}; (3) \citet{romani2005};
(4) \citet{feast1999}; (5) \citet{trimble1971}; (6) \citet{tian2008};
(7) \citet{roberts1993}; (8) \citet{kothes2001}; (9) \citet{gaensler2003};
(10) \citet{koo2008}; (11) \citet{caraveo2001}; (12) \citet{romani2005};
(13) \citet{cordes2002}
}
\tablenotetext{a}{We used parameters for outer torus.}
\end{deluxetable}

\begin{deluxetable}{p{5pc}ccc}
\tabletypesize{\scriptsize}
\tablecaption{$\sigma_0$ vs.
the expected correlation factor
and the probability of the required correlation factor
{\bf with eq.(\ref{eq:result2})}
\label{tab:fluctuation}}
\tablewidth{0pt}
\tablehead{
\colhead{$\sigma_0$} & \colhead{$\sigma_0$/2} & \colhead{$<A>$} & \colhead{$P(A>0.82)$}
}
\startdata
0.4\dotfill & 0.2 & 0.90 & 0.93 \\
0.56\dotfill & 0.28 & 0.82 & 0.56 \\
0.6\dotfill & 0.3 & 0.80 & 0.49 \\
0.8\dotfill & 0.4 & 0.71 & 0.23 \\
1.0\dotfill & 0.5 & 0.62 & 0.12
\enddata
\end{deluxetable}

\begin{deluxetable}{p{5pc}ccc}
\tabletypesize{\scriptsize}
\tablecaption{$\sigma_0$ vs.
the expected correlation factor
and the probability of the required correlation factor
with eq.(\ref{eq:9samples})
\label{tab:fluctuation2}}
\tablewidth{0pt}
\tablehead{
\colhead{$\sigma_0$} & \colhead{$\sigma_0$/2} & \colhead{$<A>$} & \colhead{$P(A>0.93)$}
}
\startdata
0.2\dotfill & 0.1 & 0.98 & 1.00 \\
0.3\dotfill & 0.15 & 0.95 & 0.79 \\
0.36\dotfill & 0.18 & 0.93 & 0.56 \\
0.4\dotfill & 0.2 & 0.91 & 0.42 \\
0.6\dotfill & 0.3 & 0.83 & 0.11 \\
0.8\dotfill & 0.4 & 0.74 & 0.04 
\enddata
\end{deluxetable}

\end{document}